\documentclass[aps,pra,twocolumn,superscriptaddress,nofootinbib]{revtex4}

\usepackage{graphicx}	
\usepackage{amssymb}	
\usepackage{amsfonts}	
\usepackage{amsmath}	
\usepackage{xcolor}		
\usepackage[T1]{fontenc}	
\usepackage[utf8]{inputenc} 
\usepackage{makecell} 
\usepackage{booktabs} 

\usepackage[english]{babel} 	
\usepackage[colorlinks=true,linkcolor=blue,citecolor=blue]{hyperref} 

\usepackage[caption=false]{subfig}		
\usepackage{braket}		
\usepackage{bm}		
\usepackage[load-configurations = abbreviations]{siunitx}	
\newcommand{\matr}[1]{\mathbf{#1}}  

\begin{document}

\title{Perturbation theory and thermal transport in mass-disordered alloys:  Insights from Green’s function methods}

\author{S. Th\'ebaud}
\email[E-mail: ]{thebaudsj@ornl.gov}
\affiliation{Materials Science and Technology Division, Oak Ridge National Laboratory, Oak Ridge, Tennessee 37831, USA}
\author{T. Berlijn}
\affiliation{Center for Nanophase Materials Sciences, Oak Ridge National Laboratory, Oak Ridge, Tennessee 37831, USA}
\affiliation{Computational Sciences and Engineering Division, Oak Ridge National Laboratory, Oak Ridge, Tennessee 37831, USA}
\author{L. Lindsay}
\email[E-mail: ]{lindsaylr@ornl.gov}
\affiliation{Materials Science and Technology Division, Oak Ridge National Laboratory, Oak Ridge, Tennessee 37831, USA}

\begin{abstract}
Lowest-order quantum perturbation theory (Fermi’s golden rule) for phonon-disorder scattering has been used to predict thermal conductivities in several semiconducting alloys with surprising success given its underlying hypothesis of weak and dilute disorder. In this paper, we explain how this is possible by focusing on the case of maximally mass-disordered Mg$_2$Si$_{1-x}$Sn$_x$.  We use a Chebyshev polynomials Green’s function method that allows a full treatment of disorder on very large systems (tens of millions of atoms) to probe individual phonon linewidths and frequency-resolved thermal transport. We demonstrate that the success of perturbation theory originates from the specific form of mass disorder terms in the phonon Green’s function and from the interplay between anharmonic and disorder scattering. 
\end{abstract}

\maketitle

\section{Introduction}

Vibrational thermal properties of condensed matter systems play critical roles in determining material stability and synthesizability, properties and functionalities, and resulting pathways towards building useful multicomponent, multiscale devices for a variety of technologies \cite{champierThermoelectricGeneratorsReview2017,bellCoolingHeatingGenerating2008,mooreEmergingChallengesMaterials2014,shinAdvancedMaterialsHighTemperature2019}. Often material synthesis pathways are designed to reduce crystal imperfections for improved properties such as low electrical and thermal resistances. However, sometimes defects and disorder are intentionally created to improve structural and elastic behaviors or increase thermal resistance \cite{horiTuningPhononTransport2018,beekmanInorganicCrystalsGlass2017,hanusThermalTransportDefective2021}. Alloys represent such a class of materials which have played critically important roles in technological advancements over the course of human history, e.g., bronze tools and weapons, steel construction materials, and high temperature aluminum alloys in combustion engines. 

In terms of thermal management applications, a variety of novel alloys have been developed and employed, particularly towards reducing vibrational contributions to thermal transport for more efficient thermoelectrics – example alloys include Si$_{1-x}$Ge$_x$ \cite{basuHighTemperatureSi2021}, PbTe$_{1-x}$Se$_x$ \cite{lalondeLeadTellurideAlloy2011}, half-Heusler alloys \cite{huangRecentProgressHalfHeusler2016}, and Mg$_2$Si$_{1-x}$Sn$_x$ \cite{pandelReviewMg2Si2021,farahiHighEfficiencyMg22019}. To build insights into the vibrational behaviors of these materials, density functional theory (DFT) based Peierls-Boltzmann transport models have been developed and deployed to examine intrinsic and extrinsic phonon scatterings and resulting thermal conductivities ($\kappa$) as a function of alloy concentration ($x$) \cite{gargRoleDisorderAnharmonicity2011,liThermalConductivityBulk2012,tianPhononConductionPbSe2012}. These models have been surprisingly successful despite the array of \textit{ad hoc} approximations employed: virtual crystal approximation (VCA), perturbation theory for mass disorder (Fermi’s golden rule), and lack of force disorder. In particular, the VCA ignores short range variations in the alloy systems and perturbation theory is not well-justified for systems with strong disorder, as is the case for the alloys described above. A case may be made for ignoring force disorder as the alloying elements are often isoelectronic; however, more recent DFT Green’s function calculations of phonon-disorder scattering and $\kappa$ in In$_{1-x}$Ga$_x$As have demonstrated that even this assumption is not fully valid \cite{arrigoniFirstprinciplesQuantitativePrediction2018b}. 
     
Building on previous work featuring mass disorder in spring-mass models \cite{thebaudSuccessBreakdownTmatrix2020}, here we critically examine the widely used VCA and perturbation approximations for describing phonon-mass-disorder scattering in alloyed semiconductors from DFT methods with specific application to maximally mass disordered Mg$_2$Si$_{0.5}$Sn$_{0.5}$. We confront this standard methodology with a more rigorous non-perturbative approach, the Chebyshev polynomials Green’s function method (CPGF) \cite{thebaudSuccessBreakdownTmatrix2020,bouzerarDrasticEffectsVacancies2020,muUnfoldingComplexityPhonon2020,allenRecoveringHiddenBloch2013}, thus isolating the effect of the perturbative approximation on the phonon lifetimes and on the thermal conductivity. We find that the general VCA phonon quasiparticle picture breaks down above a certain frequency but succeeds in describing low-frequency long-wavelength acoustic phonons that carry much of the heat in this system.  Moreover, the presence of an ordered Mg sublattice reduces the sensitivity to disorder of dispersive optic phonons, enabling them to play a significant role in heat transport. In light of these results, we discuss the conditions under which we expect the thermal conductivity to be well-predicted by perturbation theory.

\section{Mass disorder models}

In this section we describe calculations of the phonon spectrum and phonon transport in mass-disordered alloys from (i) the virtual crystal approximation and Fermi’s golden rule (quantum perturbation theory) and (ii) the Chebyshev polynomials Green’s function method \cite{ferreiraCriticalDelocalizationChiral2015}. 

\subsection{The virtual crystal approximation and Fermi’s golden rule}

In the VCA, the various properties of the alloy (e.g., masses, lattice constants, interatomic forces) are determined by the weighted average of the same properties of the end member materials based on the alloy concentration \cite{abelesLatticeThermalConductivity1963}. That is, the alloy is treated as a perfect crystal with modified phonon dispersion and other intrinsic properties based on the end members. The phonon scattering rates due to alloy mass disorder are then calculated via quantum perturbation theory, i.e., Fermi’s golden rule (FGR) \cite{tamuraIsotopeScatteringDispersive1983,tamuraIsotopeScatteringLargewavevector1984}:
\begin{equation}
\frac{1}{\tau^\text{FGR}_{\bm{q}j}} = \frac{\pi}{2 N} \omega^2_{\bm{q}j} \sum_k g_k \sum_{\bm{q}'j'} |\bm{e}_{k,\bm{q}j} \cdot \bm{e}^*_{k,\bm{q}'j'} |^2 \delta (\omega_{\bm{q}j} - \omega_{\bm{q}'j'})
\end{equation}
with $N$ the number of cells in the crystal, $\omega_{\bm{q}j}$ the VCA frequency for phonon with wavevector $\bm{q}$ and polarization $j$, and $\bm{e}_{k,\bm{q}j}$ the normalized VCA eigenvector for the atomic site that the alloyed atoms occupy labeled by $k$. Here, $g_k = f_k (m_k - m_\text{vc})^2/m_\text{vc}^2$ is a mass variance parameter for the $k^\text{th}$ atom type of the alloyed site with $m_k$ being the mass of the $k^\text{th}$ atom type, $m_\text{vc}$ being the VCA averaged mass, and $f_k$ being the concentration of the $k^\text{th}$ atom in the alloy. The total phonon scattering rate can be obtained from Matthiessen’s rule: $\frac{1}{\tau_{\bm{q}j}} = \frac{1}{\tau^\text{FGR}_{\bm{q}j}} + \frac{1}{\tau^\text{3ph}_{\bm{q}j}}$, where $\tau^\text{3ph}_{\bm{q}j}$ is the lifetime limited by anharmonicity, most often computed from 3-phonon interactions. Within the relaxation time approximation, the thermal conductivity ($\kappa$) along a particular direction $x$ can be determined by:
\begin{equation}
\kappa = \frac{1}{\Omega} \sum_{\bm{q}j} C(\omega_{\bm{q}j}) v_{x,\bm{q}j}^2 \tau_{\bm{q}j}
\end{equation}
with $\Omega$ the volume, $C(\omega_{\bm{q}j})$ the heat capacity and $v_{x,\bm{q}j}$ the velocity of the mode in the $x$ direction.

This methodology has been widely used in the literature to predict the thermal conductivity of various semiconducting alloys \cite{tianPhononConductionPbSe2012,lindsayCalculatedTransportProperties2015,pandeyInitioPhononThermal2017,leeLatticeThermalConductivity2014,maIntrinsicThermalConductivities2016}.  
In some studies, the phonon-disorder scattering has been treated using the T-matrix approximation, which takes into account multiple scatterings off a single impurity and is therefore exact in the limit of dilute disorder \cite{kunduRoleLightHeavy2011a,arrigoniFirstprinciplesQuantitativePrediction2018b}. However, for mass-disordered binary alloys at the $50\%$ composition, multiple-occupancy corrections to the T-matrix cancel all contributions from three or more scatterings, leaving only the FGR term (see eq.~(20) of Ref.~\onlinecite{schwartzComparisonAveragetMatrixCoherentPotential1971} and Fig.~9 of Ref.~\onlinecite{elliottTheoryPropertiesRandomly1974}). Thus, we only consider the FGR in the present study. In particular, it has been very successful for Si$_{1-x}$Ge$_x$ \cite{gargRoleDisorderAnharmonicity2011} and Mg$_2$Si$_{1-x}$Sn$_x$ \cite{liThermalConductivityBulk2012}, two cases when a straightforward comparison with experiment has been possible. This success is unexpected since the FGR, as a lowest-order perturbation theory, is supposed to be valid only for weak, dilute disorder. In Si$_{1-x}$Ge$_x$ with maximal disorder ($x = 0.5$), for instance, every atom can be considered a defect, and the mass ratio between Ge and Si is close to 2.5. Therefore, the disorder is neither weak nor dilute and the success of the FGR is confounding.

\subsection{Chebyshev polynomials Green’s function method}

In order to shed light on this puzzle, we confront the FGR with another approach based on the formalism of Green’s functions, the CPGF method, which does not require assumptions of weak or dilute disorder. Indeed, it is a nonperturbative technique that allows for a full treatment of disorder in very large systems. In this way, we can directly evaluate the validity or breakdown of lowest-order perturbation theory by comparing phonon lifetimes and resulting conductivities predicted by the two methods. Moreover, comparisons can be made with frequency resolution and without having to disentangle the effects of anharmonicity, boundary scattering, modeling parameters, and measurement uncertainties as found in experiments. 

The CPGF method \cite{ferreiraCriticalDelocalizationChiral2015} relies on a real-space representation of the phonon Green’s function of a large, disordered supercell (see Appendix~\ref{appendix_green_function}). The retarded Green’s function $\matr{G}(\omega)$ is a frequency-dependent operator that can be defined \cite{mingoClusterScatteringEffects2010,elliottTheoryPropertiesRandomly1974} as
\begin{equation}
\matr{G}(\omega) = \frac{1}{(\omega + i \eta)^2 - \matr{D}}
\end{equation}
in which $\eta$ is a real positive infinitesimal and $\matr{D} = \frac{1}{\sqrt{\matr{M}}} \matr{\Phi} \frac{1}{\sqrt{\matr{M}}}$ is the dynamical matrix of the supercell with $\matr{M}$ the diagonal matrix of the masses and $\matr{\Phi}$ the matrix of the force constants. Since we neglect force constant disorder here, $\matr{\Phi}$ is ordered while $\matr{M}$ and $\matr{D}$ are disordered. Evaluation of the Green’s function allows determination of the phonon density of states (DOS) and spectral function (see Appendix~\ref{appendix_green_function}). In the CPGF method, the phonon Green’s function is expanded on the basis of Chebyshev polynomials, with an efficient iterative evaluation of the successive terms in the expansion (see Appendix~\ref{appendix_CPGF}). Like exact diagonalization, this is a full treatment of disorder in the sense that all diagrams in the self-energy expansion are incorporated, including vertex corrections for two-particle quantities such as the thermal conductivity (see below). Unlike exact diagonalization, the favorable $O(N)$ scaling of CPGF allows for very large supercell sizes of tens of millions of atoms. For such systems, one disorder configuration is sufficient to obtain self-averaged properties. However, the method also scales linearly with the number of harmonic interatomic force constants (IFCs) per atom, rendering it impractical when long-range force constants have to be included.

Phonon-disorder scattering rates can be extracted from the full width at half-maximum of peaks in the spectral function for given VCA phonon modes, in the same manner as in Ref.~\onlinecite{thebaudSuccessBreakdownTmatrix2020} and Ref.~\onlinecite{bouzerarDrasticEffectsVacancies2020}. Large system sizes provide sufficient resolution to probe the lifetimes of low-frequency acoustic modes. Frequency-resolved thermal transport can also be calculated directly using the Kubo formalism (see Ref.~\onlinecite{flickerLatticeThermalConductivity1973,allenThermalConductivityDisordered1993} and Appendix~\ref{appendix_green_function}), which does not assume the presence of well-defined phonon quasiparticles and does not require phonon-disorder lifetimes as inputs. The thermal conductivity in the $x$ direction can be written as 
\begin{equation}
\kappa = \int_0^\infty d\omega \, W_\text{ph} (\omega) \Sigma_\text{ph} (\omega) 
\end{equation}
with $ W_\text{ph} (\omega)$ a normalized $k_B T$ window (see Appendix~\ref{appendix_green_function}) and the phonon transport distribution function (TDF) $ \Sigma_\text{ph} (\omega)$ is:
\begin{align}
\label{kubo}
\Sigma_\text{ph} (\omega) = \frac{\pi k_B ^2 T}{3 \hbar \Omega} \text{Tr} \bigg[ & \text{Im} \matr{G}(\omega + i \frac{\Gamma_\text{3ph}(\omega)}{2}) \matr{S}_x \; \\ 
& \times \text{Im} \matr{G}(\omega + i \frac{\Gamma_\text{3ph}(\omega)}{2}) \matr{S}_x \bigg] \nonumber
\end{align}
where $T$ is the temperature, $\Omega$ is the supercell volume and $\matr{S}_x$ is the heat current operator in the $x$ direction. The trace over the supercell degrees of freedom is computed via an efficient stochastic method (see Appendix~\ref{appendix_CPGF}). A frequency-dependent phonon-phonon scattering rate $\Gamma_\text{3ph}$ is included as an imaginary part in the Green’s function to account for phonon-phonon interactions in a simple way. 

It is informative to compare our present approach to thermal conductivity calculations with theories beyond Peierls-Boltzmann as recently proposed by Simoncelli, Caldarelli, and Isaeva. Simoncelli’s paper establishes a Wigner formalism for complex crystals that separates two terms in the thermal conductivity: diagonal propagative transport described by Boltzmann-Peierls theory and off-diagonal diffusive transport involving phonon modes mixing via mode-dependent broadening due to anharmonicity and disorder \cite{simoncelliUnifiedTheoryThermal2019}. In addition, a very recent paper by Caldarelli extends this formalism to the overdamped regime by using the Green-Kubo formula on the basis of the phonon modes and, neglecting vertex corrections (dressed bubble approximation), writing it in terms of the phonon spectral function \cite{caldarelliManybodyGreenFunction2022}. Isaeva’s approach to thermal transport in disordered systems uses the Green-Kubo formula expressed on the basis of disordered normal modes of the supercell and introduces a constant anharmonic lifetime in the Green’s function for each normal mode \cite{isaevaModelingHeatTransport2019}. Our approach is essentially equivalent to Isaeva’s, but the Green-Kubo formula is implemented in real-space (not using the normal modes) and the anharmonicity is included through a frequency-dependent lifetime, allowing us to study very large systems (tens of millions of atoms) through the CPGF algorithm. In Caldarelli’s language, disorder is treated at the ‘FSF’ level (actually beyond FSF, because vertex corrections are included) and anharmonicity at the ‘LSFA’ level  \cite{caldarelliManybodyGreenFunction2022}.  Thus, our approach can describe systems simultaneously featuring arbitrarily strong disorder (phonons overdamped by disorder, including localization effects) and wave-like tunneling through anharmonic overlap of normal modes.

\subsection{DFT calculations: Mg$_2$Si$_{0.5}$Sn$_{0.5}$}

\begin{figure}
\centering
\includegraphics[width=0.8\columnwidth]{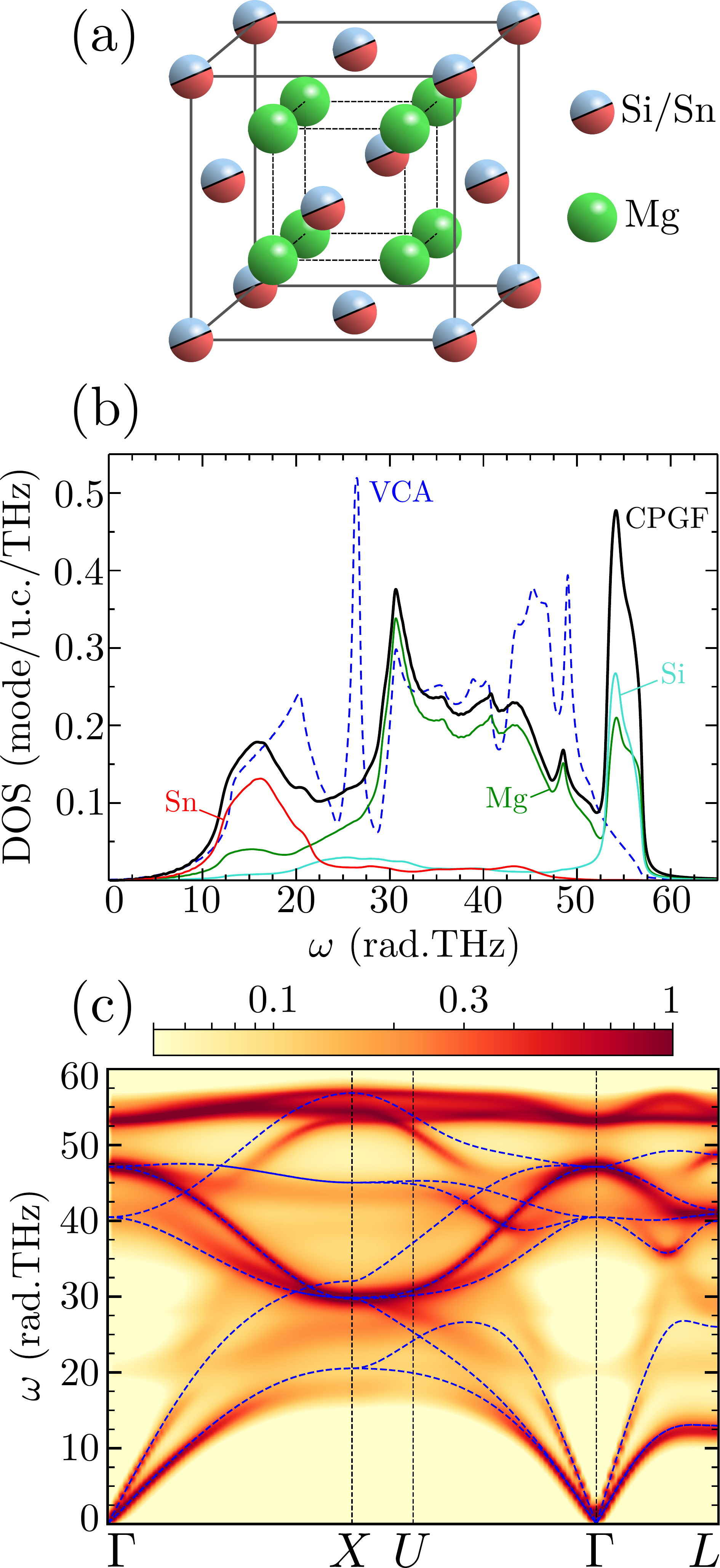}
\caption{(a) Atomic structure of Mg$_2$Si$_{0.5}$Sn$_{0.5}$. (b) Calculated phonon density of states (DOS) from the virtual crystal approximation (VCA, dashed blue curve) and from the Chebyshev polynomials Green’s function method (CPGF, solid black curve). The DOS projected on Mg, Si, and Sn atoms are also shown (green, cyan, and red curves, respectively). (c) Phonon spectral function calculated by the CPGF method and the VCA dispersion (dashed blue curves). The intensity scale is in \si{(rad.THz)^{-1}}.}
\label{fig1}
\end{figure}

We focus our study to the case of maximally disordered Mg$_2$Si$_{0.5}$Sn$_{0.5}$ as a representative of the class of semiconducting alloys that are well-described by the FGR. It crystallizes in the antifluorite structure (see Fig.~\ref{fig1}(a)) with Si (\SI{28.09}{amu}) and Sn (\SI{118.71}{amu}) atoms (mass ratio $\sim 4$) randomly placed in a disordered fcc sublattice. The Mg (\SI{24.31}{amu}) atoms occupy the tetrahedral sites, forming an ordered sublattice. 

The second-order IFCs of the end members Mg$_2$Si and Mg$_2$Sn were computed from density functional perturbation theory \cite{baroniPhononsRelatedCrystal2001}. To make the CPGF calculations tractable, we neglected the long-range Coulomb corrections and applied a \SI{6}{\angstrom} cutoff for atomic interactions before enforcing the acoustic sum rule. The frequency-dependent three-phonon scattering rates $\Gamma_\text{3ph}(\omega) = 1/\tau^\text{3ph}(\omega)$ were determined by averaging the mode-dependent lifetimes $\tau^\text{3ph}_{\bm{q}j}$ over phonon wavevectors and branches.  The $\tau^\text{3ph}_{\bm{q}j}$ were built from harmonic and third-order IFCs previously obtained in Ref.~\onlinecite{liThermalConductivityBulk2012}. Further details can be found in Appendix~\ref{appendix_mg2sisn_model}.

The aforementioned averaging procedure and truncated range of the harmonic IFCs are approximations made so that numerical implementation of the CPGF method is tractable. This introduces some deviation from measured vibrational frequencies and lattice thermal conductivities, though still allows for direct comparison of the FGR and CPGF methods, as the approximations are made in both cases.

\section{Results and discussion}

Fig.~\ref{fig1}(b) gives the total and projected phonon density of states (DOS) of Mg$_2$Si$_{0.5}$Sn$_{0.5}$ calculated with the CPGF method and compared with that from the VCA, while Fig.~\ref{fig1}(c) gives the CPGF phonon spectral function along high-symmetry lines compared with the VCA calculated dispersion. There are significant differences between the VCA and CPGF spectra. Most prominently, there is a frequency range (from \SI{15}{rad.Thz} to \SI{30}{rad.THz}) corresponding to the top of the acoustic branches where the phonon modes are so broadened by disorder that they can no longer be described as well-defined quasiparticles and have a diffuson character in Allen and Feldman's terminology \cite{allenDiffusonsLoconsPropagons1999}. As a result, the lowest Van Hove (VH) singularity at \SI{13}{rad.THz} is partially smoothed out, while the higher VH singularity at \SI{26}{rad.THz} is entirely destroyed. The CPGF spectrum also features two flat branches at the high end of the spectrum (\SI{55}{rad.THz}) that are absent in the  VCA. The vibrational character of these branches is almost entirely on the light atoms Mg and Si, which is consistent with their high frequency. The other optic modes, from \SI{30}{ rad.THz} to \SI{50}{rad.THz}, are dominated by Mg vibrations. Since the Mg sublattice is not disordered, many of these modes are not sensitive to disorder. Indeed, one CPGF branch is identical to its VCA counterpart on the $\Gamma-X-U-\Gamma$ high symmetry path, though not so along the $\Gamma-L$ segment. Not surprisingly, the mode broadenings become vanishingly small close to the $\Gamma$ point as they are protected by translation invariance.  

Figure 2 gives the phonon-disorder contribution to the inverse lifetimes $\Gamma_\text{dis}$ for the acoustic modes from separate FGR and CPGF calculations. These are also compared with the calculated anharmonic scattering rates $\Gamma_\text{3ph}(\omega)$ at \SI{300}{K} in the same frequency range. Both the FGR and the CPGF method predict that $\Gamma_\text{dis}$ follows an $\omega^4$ Rayleigh power law at low frequencies, as expected from point defects \cite{klemensScatteringLowFrequencyLattice1955}. As a result, disorder scattering is smaller than anharmonic scattering for phonons below \SI{6}{rad.THz}, but dominates for acoustic phonons above this frequency.  Surprisingly, FGR and CPGF acoustic linewidths below \SI{13}{rad.THz} are in agreement; however, there is a significant discrepancy between calculated  acoustic linewidths above \SI{13}{rad.THz}: FGR linewidths are roughly 50\% larger than those from the CPGF method. As mentioned in Ref.~\onlinecite{thebaudSuccessBreakdownTmatrix2020}, the unexpected success of the FGR to describe strong disorder at low frequencies is due to the particular form of the mass perturbation term in the phonon spectral function. The Green's function enters its definition~\eqref{def_spectral_function} as 
\begin{equation}
\sqrt{\frac{\matr{M}_\text{vc}}{\matr{M}}} \matr{G}(\omega) \sqrt{\frac{\matr{M}_\text{vc}}{\matr{M}}} = \frac{1}{(\omega + i \eta)^2 - \matr{D}_\text{vc} + \frac{\Delta \matr{M}}{\matr{M}_\text{vc}} \omega^2}
\end{equation}
where $\matr{D}_\text{vc} = \frac{1}{\sqrt{\matr{M}_\text{vc}}} \matr{\Phi} \frac{1}{\sqrt{\matr{M}_\text{vc}}}$ is the VCA dynamical matrix and $\Delta \matr{M} = \matr{M} - \matr{M}_\text{vc}$ is the mass perturbation. The perturbation term is itself proportional to $\omega^2$, therefore this type of disorder is weak at low frequencies. By contrast, an IFC perturbation term would be of the form $\frac{1}{\sqrt{\matr{M}_\text{vc}}} \Delta \matr{\Phi} \frac{1}{\sqrt{\matr{M}_\text{vc}}}$, in which the frequency does not appear explicitly. As a consequence, it is possible to estimate at which frequency the FGR for mass disorder can be expected to fail by a straightforward evaluation of the dominant higher order term in the standard perturbative expansion of the phonon self-energy (see Appendix~\ref{appendix_diagrams} for more details). In the specific case of Mg$_2$Si$_{0.5}$Sn$_{0.5}$, the breakdown of the FGR occurs around the frequency of the lowest VH singularity at \SI{13}{rad.THz}. The sharp upturn in the DOS is fully reflected in the FGR scattering rates, but has much less of an impact on the CPGF lifetimes, leading to an overestimation of the scattering by the FGR.

\begin{figure}
\centering
\includegraphics[width=1.0\columnwidth]{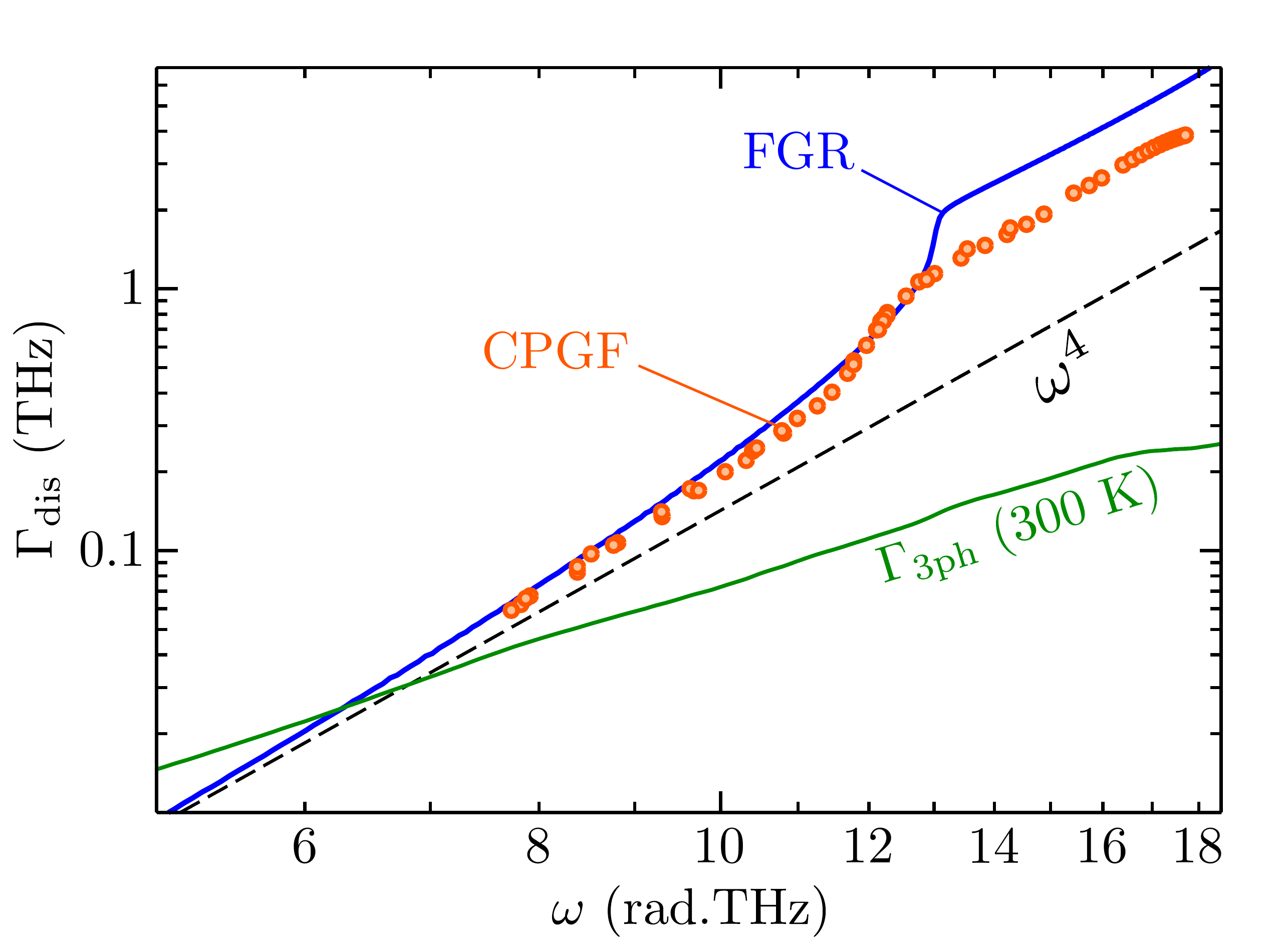}
\caption{Acoustic phonon linewidths as a function of angular frequency calculated via CPGF (orange circles) and FGR (blue curve). Room temperature anharmonic phonon-phonon scattering rates are also shown (green curve).}
\label{fig2}
\end{figure}

\begin{figure}
\centering
\includegraphics[width=1.0\columnwidth]{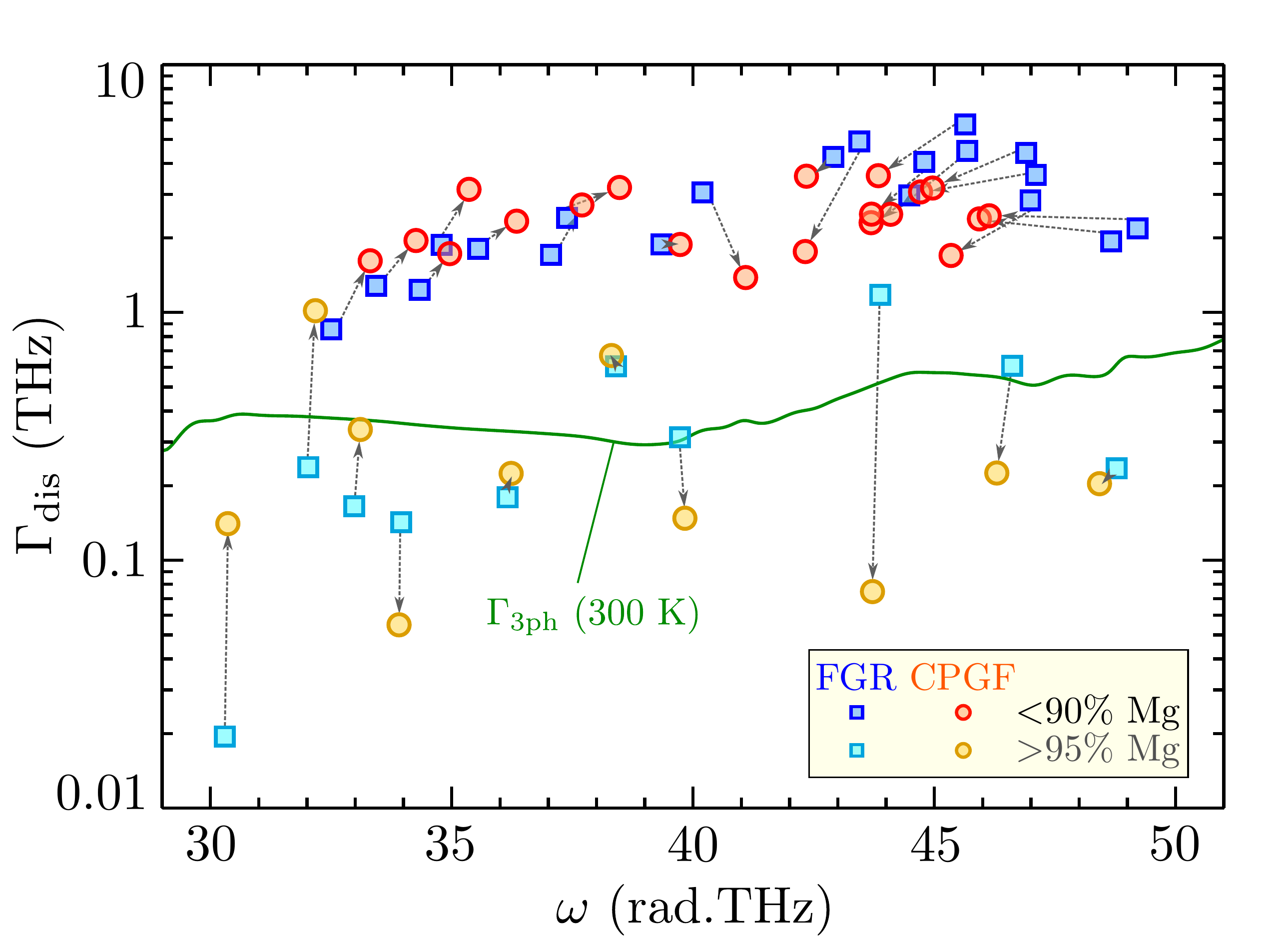}
\caption{Phonon linewidths for selected optic modes calculated by the CPGF method as a function of the renormalized mode angular frequency (circles) and calculated by Fermi’s golden rule (FGR) as a function of the VCA angular frequency (squares). Cyan and orange data correspond to modes with a high Mg vibrational character ($> 95\%$), while blue and red data correspond to modes with lower Mg vibrational character ($ < 90\%$). The room temperature anharmonic phonon-phonon scattering rates are also shown (green curve).}
\label{fig3}
\end{figure}

As will be seen below, the optic phonons contribute significantly to the thermal conductivity of Mg$_2$Si$_{0.5}$Sn$_{0.5}$, so it is worth examining their scattering. Figure \ref{fig3} gives the disorder-induced inverse lifetimes of selected optic phonon modes between \SI{30}{rad.THz} and \SI{50}{rad.THz}, along with the room-temperature anharmonic scattering rates. The $x$-axis corresponds to VCA frequencies for the FGR data and to disorder-renormalized frequencies (shift of the phonon mode due to the presence of disorder) for the CPGF data. The modes with high Mg vibrational weight (>95\%) predominantly sit on the ordered Mg sublattice, thus their disorder scattering rate is comparable to or weaker than the anharmonic scattering rates and have very small frequency renormalization. By contrast, modes with a lower Mg vibrational character have significant weight on the Si/Sn sublattice, so phonon-disorder scattering dominates over anharmonic scattering and the frequencies are substantially renormalized. For most phonons, including many modes with high Mg vibrational character, the FGR breaks down, resulting in large errors compared to the CPGF method. We note that the errors are irregular: the FGR tends to underestimate $\Gamma_\text{dis}$ in the \SIrange{30}{40}{rad.THz} range (though not always), while it tends to overestimate $\Gamma_\text{dis}$ in the \SIrange{40}{50}{rad.THz} range.

\begin{figure}
\centering
\includegraphics[width=1.0\columnwidth]{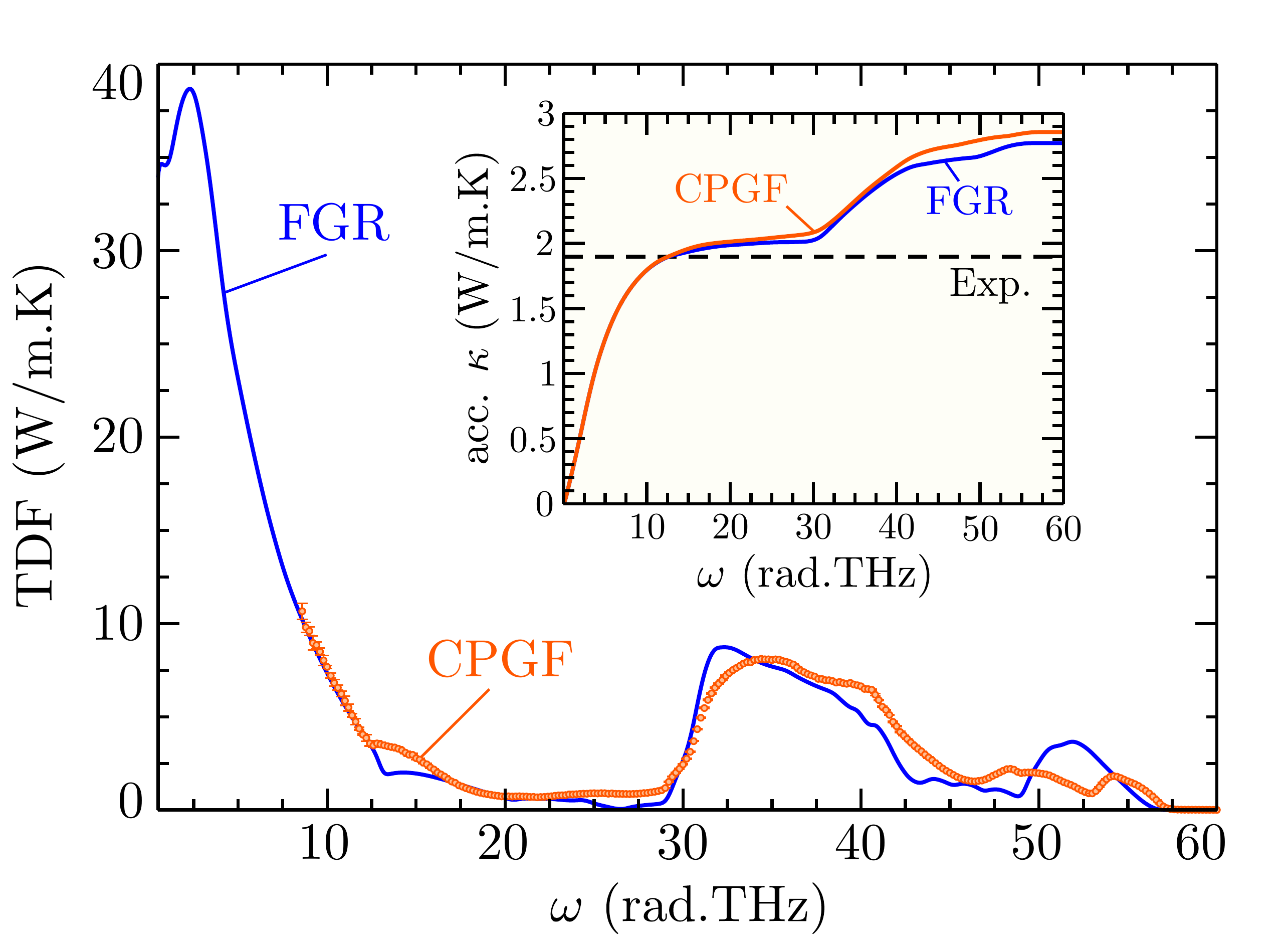}
\caption{Room temperature phonon transport distribution function (TDF) calculated by the Boltzmann transport equation under the relaxation time approximation with the FGR phonon-disorder scattering rates (blue curve) and calculated by the Green-Kubo formalism with the CPGF method (orange circles). The error bars on the circles give an estimate of the uncertainty associated with the stochastic evaluation of the trace (see Appendix~\ref{appendix_CPGF}). The inset gives the corresponding accumulated thermal conductivities. The total measured value from Ref.~\onlinecite{liThermalConductivityBulk2012} is also shown (dashed black line).}
\label{fig4}
\end{figure}

Next, we investigate how these results propagate into calculations of the thermal conductivity. Figure~\ref{fig4} gives the phonon TDF at \SI{300}{K} and the corresponding accumulated thermal conductivity (inset). The phonon TDF gives the spectral thermal conductivity contributions, which when integrated give the overall thermal conductivity. The curve labeled ‘FGR’ was calculated using the FGR phonon-disorder scattering rates and the Peierls-Boltzmann transport equation within the relaxation time approximation. The curve labeled ‘CPGF’ was calculated using the Kubo formalism (eq.~\eqref{kubo}) directly, without reliance on a phonon quasiparticle picture or the results of Figs.~\ref{fig2} and \ref{fig3}. Note that the FGR and CPGF calculated room temperature thermal conductivities presented here are significantly higher than measurements (Fig.~\ref{fig4} inset). This is partly due to approximations made for numerically tractable CPGF calculations: a range cutoff on the force constants, not including long-range polar corrections, and the averaging of the mode-dependent phonon lifetimes. In the future, this difficulty might be overcome by a modified mixed-space CPGF algorithm in which the VCA part of the supercell dynamical matrix is represented in reciprocal space.

As expected from the spectral function behavior and the acoustic phonon lifetimes, both methods agree below \SI{13}{rad.THz} and disagree between \SI{13}{rad.THz} and \SI{30}{rad.THz}. In particular, the FGR predicted TDF is much lower around \SI{26}{rad.THz} due to the enhanced scattering from the VH singularity in the VCA DOS. More surprisingly, the FGR predicted TDF is quite close to that predicted by the CPGF method for the optic modes (between \SI{30}{rad.THz} and \SI{60}{rad.THz}), which account for almost a third of the total thermal conductivity. This can be explained as follows. First, the FGR $\Gamma_\text{dis}$ is sometimes overestimated and sometimes underestimated in this frequency regime, which leads to some overall compensation. Second, the optic modes with mostly Mg vibrational character are much less scattered by disorder than modes with a higher Si/Sn vibrational character. As a result, modes with high Mg character contribute much more to the thermal conductivity (see Appendix~\ref{appendix_mg2sisn_model}). However, the phonon-disorder linewidths $\Gamma_\text{dis}$ are small relative to anharmonic scattering ($\Gamma_\text{tot} = \Gamma_\text{dis} + \Gamma_\text{3ph}$) for these heat-carrying optic modes with high Mg vibrational character. Thus, the failure of the FGR to predict $\Gamma_\text{dis}$ has little impact on determining their thermal conductivity contributions, yielding a generally accurate prediction of the TDF. 

Given the previous discussion, we expect the FGR to accurately describe thermal transport in mass-disordered alloys when optic modes have only a minor contribution to heat transport. This may be the case in compounds without compositionally distinct sublattices, such as in Si$_{1-x}$Ge$_x$, compounds with an ordered heavy sublattice supporting low-frequency acoustic modes with long mean free paths, or in situations where the optic modes are not thermally populated as occurs at temperatures significantly lower than the upper limit of the phonon spectrum. As in Mg$_2$Si$_{1-x}$Sn$_x$, we expect the FGR to yield good predictions of the thermal conductivity in mass-disordered alloys where there are disorder-insensitive optic modes sitting on an ordered light atom sublattice. However, perturbation theory may fail in compounds possessing both a disordered heavy sublattice (to suppress the conduction of acoustic modes) and a disordered light sublattice (to make disorder relevant at high frequencies). Perhaps more importantly, the FGR is expected to break down with significant IFC disorder in the alloy. Notably, it mostly fails to describe the measured thermal conductivity of III-V alloys (In,Ga)As, (Al,Ga)N, (In,Ga)N, and (In,Al)N \cite{arrigoniFirstprinciplesQuantitativePrediction2018b,maIntrinsicThermalConductivities2016}. 

\section{Summary}

To conclude, we explained the surprising success of lowest order quantum perturbation theory (i.e., Fermi’s golden rule) in describing phonon-disorder scattering and thermal conductivities in predominantly mass-disordered alloys. By confronting this standard computational approach with the non-perturbative Chebyshev polynomials Green’s function method, we demonstrated that the phonon quasiparticle picture breaks down above a certain frequency, but Fermi’s golden rule is valid for low frequency heat-carrying acoustic modes due to the $\frac{\Delta \matr{M}}{\matr{M}_\text{vc}} \omega^2$ form of the mass perturbation. In our reference system Mg$_2$Si$_{0.5}$Sn$_{0.5}$, there is a substantial contribution to the thermal conductivity from optic phonons with high Mg vibrational character, though these modes are scattered more significantly by anharmonicity at room temperature, thus masking the failure of Fermi’s golden rule for these modes. We discussed the conditions for validity of the perturbative predictions of thermal transport. In particular, we expect the perturbative approach to succeed in most mass-disordered alloys, but to fail for alloys with significant force constant disorder and in systems with significant optic mode thermal conductivity contributions derived from a disordered light atom sublattice. 

\section*{ACKNOWLEDGEMENTS}

This research was supported by the U.S. Department of Energy, Office of Science, Basic Energy Sciences, Materials Sciences and Engineering Division. We used resources of the Compute and Data Environment for Science (CADES) at the Oak Ridge National Laboratory, which is supported by the Office of Science of the U.S. Department of Energy under Contract No. DE-AC05-00OR22725.

\appendix

\section{Details of the first-principles calculations and the Mg$_2$Si$_{0.5}$Sn$_{0.5}$ model}
\label{appendix_mg2sisn_model}

\begin{figure}
\centering
\includegraphics[width=1.0\columnwidth]{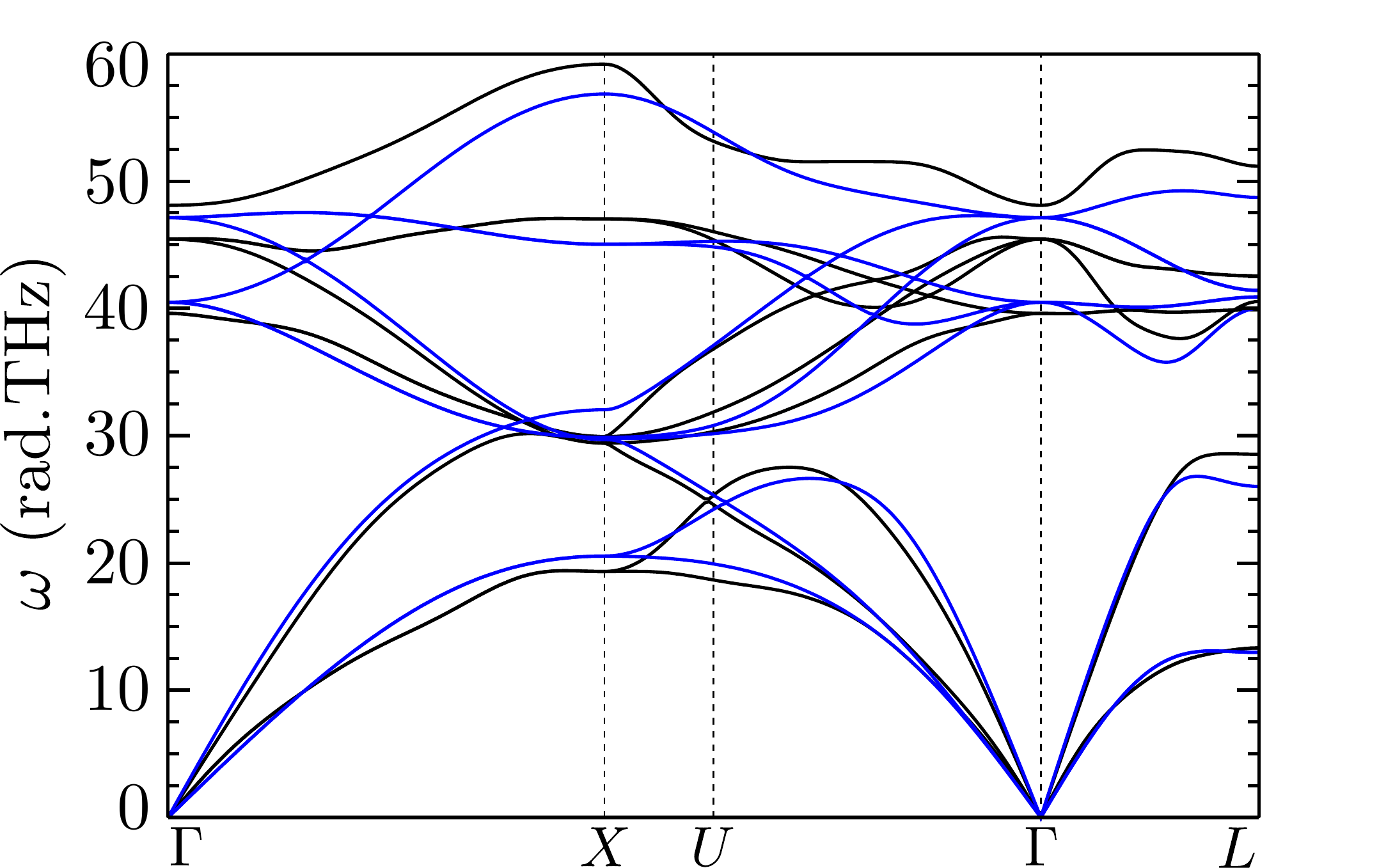}
\caption{Phonon dispersion of the virtual Mg$_2$Si$_{0.5}$Sn$_{0.5}$ crystal calculated with a full set of IFCs using an $8 \times 8 \times 8$ DFPT q-grid and long-range corrections as in Ref.~\onlinecite{liThermalConductivityBulk2012} (black curves) and with a \SI{6}{\angstrom} cutoff on the IFCs using a $3 \times 3 \times 3$ DFPT q-grid without Coulomb corrections (blue curves).}
\label{fig5}
\end{figure}

The second-order IFCs of Mg$_2$Si and Mg$_2$Sn were calculated from density functional perturbation theory (DFPT) using the Quantum Espresso package \cite{giannozziQUANTUMESPRESSOModular2009,giannozziAdvancedCapabilitiesMaterials2017} with the Perdew-Zunger LDA exchange-correlation functional \cite{perdewSelfinteractionCorrectionDensityfunctional1981}, Von Barth-Car norm-conserving pseudopotentials for Si \cite{barthCohesiveEnergyChargeDensity1985} and Bachelet-Hamann-Schlüter pseudopotentials for Mg and Sn \cite{bacheletPseudopotentialsThatWork1982}. A $16 \times 16 \times 16$ k-mesh, a wavefunction energy cutoff of \SI{70}{Ry}, and an electronic convergence threshold of \SI{e-16}{Ry} were used for the self-consistent determination of the electron density. With these parameters, the relaxed lattice constant is \SI{6.284}{\angstrom} for Mg$_2$Si and \SI{6.683}{\angstrom} for Mg$_2$Sn. A $3 \times 3 \times 3$ phonon q-grid with a tight self-consistency threshold of \SI{e-16}{Ry} was used for the DFPT calculation. In order to limit the number of IFCs, we neglected the long-range Coulomb corrections and applied a \SI{6}{\angstrom} cutoff for atomic interactions during post-processing, thus keeping only 4 neighboring atomic shells around the Si/Sn atoms and 6 shells around the Mg atoms. The acoustic sum rule was then enforced, and an average of the atomic masses and IFCs of the end members was taken to obtain a virtual Mg$_2$Si$_{0.5}$Sn$_{0.5}$ crystal (VCA). In Fig.~\ref{fig5}, we compare the VCA phonon dispersion calculated from this model with the reference dispersion computed from the full sets of IFCs used in Ref.~\onlinecite{liThermalConductivityBulk2012}. Overall, the agreement between the two is reasonable.

The VCA 3-phonon lifetimes $\tau^\text{3ph}_{\bm{q}j}$ were calculated in the standard way (eqs. (3) and (4) of Ref.~\onlinecite{liThermalConductivityBulk2012}) using the second and third-order IFCs obtained in Ref.~\onlinecite{liThermalConductivityBulk2012}. To construct a frequency-dependent quantity, we then performed the average
\begin{equation}
\frac{1}{\Gamma_\text{3ph}(\omega)} = \tau^\text{3ph}(\omega) = \frac{\sum_{\bm{q}j} \tau^\text{3ph}_{\bm{q}j} \delta(\omega - \omega_{\bm{q}j})}{\sum_{\bm{q}j} \delta(\omega - \omega_{\bm{q}j})}
\end{equation}
with the delta functions Gaussian-broadened. At low frequencies (below \SI{1}{rad.THz}) we interpolated the scattering rate to 0 by a square power law $\omega^2$. For a fair comparison between the FGR and the CPGF method, we used this frequency-dependent lifetime when calculating the TDF and thermal conductivity with both approaches. 

\begin{figure}
\centering
\includegraphics[width=1.0\columnwidth]{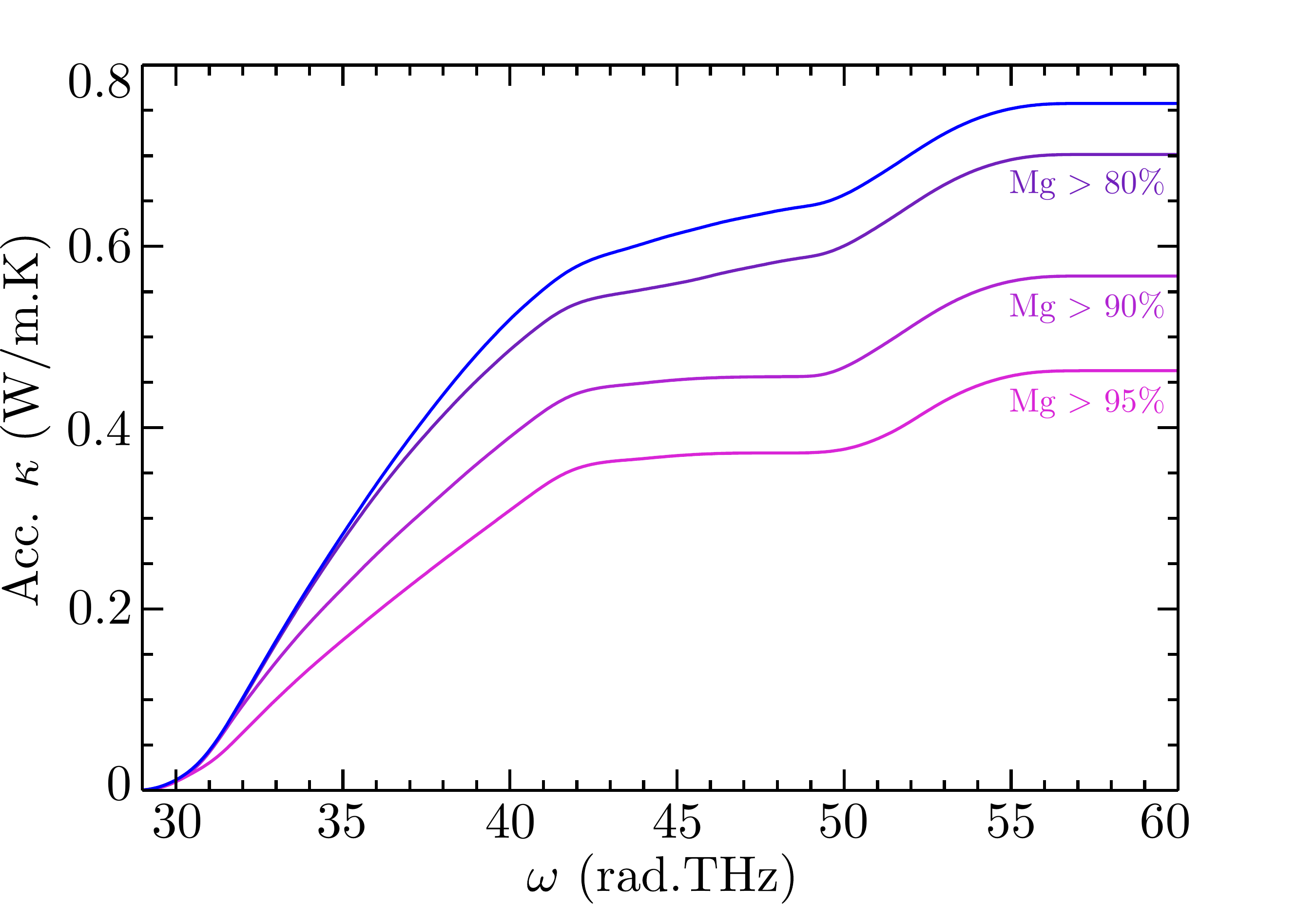}
\caption{Room temperature accumulated optic mode thermal conductivity calculated by the Boltzmann transport equation under the relaxation time approximation with the FGR phonon-disorder scattering rates. We show the contribution from all the optic modes (blue curve), and also from the optic modes with a Mg character higher than 80\%, 90 \%, and 95 \% (other curves).}
\label{fig6}
\end{figure}

In order to evaluate the contribution to thermal transport of optical modes with high Mg character, we plot in Fig.~\ref{fig6} the accumulated thermal conductivity at \SI{300}{K} calculated from the Peierls-Boltzmann transport equation with the FGR. The blue curve represents the contribution of all the optic modes while the other curves only contain the contributions from the optic modes with a Mg weight higher than 80\%, 90\%, and 95\%. Here, the Mg weight is defined as the squared norm of the projection of the mode polarisation vector onto the Mg atoms. Hence, acoustic modes near the $\Gamma$ point have a Mg weight around 40\%. For the optic modes, most of the heat is carried by modes with a high Mg weight (80\% or more), and a slight majority is carried by modes with a very high Mg weight (95\% or more). This is consistent with the fact that the Mg sublattice is ordered, and thus less sensitive to phonon-disorder scattering.

\section{Green's function formula for the phonon spectral function and thermal conductivity}
\label{appendix_green_function}

We define the phonon Green's function  on a supercell containing $N_c$ unit cells as
\begin{equation}
\matr{G}(\omega) = \frac{1}{(\omega + i \eta)^2 - \matr{D}}
\end{equation}
where $\matr{D} = \frac{1}{\sqrt{\matr{M}}} \matr{\Phi} \frac{1}{\sqrt{\matr{M}}}$ is the dynamical matrix of the supercell. $\matr{M}$ is the diagonal matrix of the atomic masses, $\matr{\Phi}$ is the matrix of the interatomic force constants and $\eta$ is a real positive infinitesimal. In the present case, the disorder on the force constants are neglected, thus only $\matr{M}$ and $\matr{D}$ are disordered. $\matr{G}(\omega)$ is the Fourier transform of the correlation function of the mass-renormalized atomic displacement operators \cite{thebaudSuccessBreakdownTmatrix2020}. This definition is consistent with most of the T-matrix literature \cite{mingoClusterScatteringEffects2010}, but different choices are sometimes made in other contexts  \cite{elliottTheoryPropertiesRandomly1974,mahanManyParticlePhysics2000}. 

The phonon density of states (DOS) $\rho(\omega)$ can be calculated from the Green's function as
\begin{equation}
\label{def_dos}
\rho(\omega) = -\frac{2 \omega}{\pi}\text{Tr} \left( \text{Im}  \matr{G}(\omega) \right)
\end{equation}
with the trace running over all degrees of freedom in the supercell (directions $\alpha$, unit cells $R$, and unit cell atoms $i$). For projections of the DOS on specific atom types, the trace becomes a partial trace on the degrees of freedom corresponding to the atom type. 

The spectral function for the VCA mode associated with wavevector $\bm{q}$ and branch $j$ can be defined (see \cite{thebaudSuccessBreakdownTmatrix2020}, eq. (B12)) as
\begin{equation}
\label{def_spectral_function}
A_j(\bm{q},\omega) = -\frac{2 \omega}{\pi}\text{Im} \bra{E_{\bm{q}j}} \sqrt{\frac{\matr{M}_\text{vc}}{\matr{M}}} \matr{G}(\omega) \sqrt{\frac{\matr{M}_\text{vc}}{\matr{M}}} \ket{E_{\bm{q}j}}
\end{equation}
where $\ket{E_{\bm{q}j}} = \frac{1}{\sqrt{N_c}} \sum_{\alpha i \bm{R}} e^{i\bm{q}(\bm{R}+\bm{r}_i)} e^\alpha_{i,\bm{q}j} \ket{\alpha i \bm{R}}$ is the eigenvector corresponding to phonon mode $(\bm{q}j)$ of the VCA dynamical matrix of the supercell. $\alpha$ runs over the directions ($x,y,z$), $\bm{R}$ over the unit cell positions and $i$ over the atoms in the unit cell. $\bm{r}_i$ is the position of atom $i$ with respect to the unit cell and $\bm{e}_{\bm{q}j}$ is the polarisation vector for mode $(\bm{q}j)$. While the present formalism is quantum mechanical \cite{thebaudSuccessBreakdownTmatrix2020}, it should be stressed that the bra-ket notation is used here strictly for convenience, as the vectors denote atomic degrees of freedom and not actual quantum states of the phonon Fock space.

In the framework of the Kubo formalism, the lattice thermal conductivity can be expressed through the Green-Kubo formula (see Ref.~\onlinecite{flickerLatticeThermalConductivity1973} or eqs. (2.50), (2.79b) and (2.83) in Ref.~\onlinecite{elliottTheoryPropertiesRandomly1974}, note the different definition for the Green's function):
\begin{align}
\label{kubo_kappa}
& \bm{\kappa} = \int_0^\infty d\omega \, W_\text{ph} (\omega) \frac{\pi k_B ^2 T}{3 \hbar \Omega} \\
 \times \text{Tr} & \left[ \text{Im} \matr{G}(\omega + i \frac{\Gamma_\text{3ph}(\omega)}{2}) \matr{S} \; \text{Im} \matr{G}(\omega + i \frac{\Gamma_\text{3ph}(\omega)}{2}) \matr{S} \right] \nonumber
\end{align}
where $T$ is the temperature, $\Omega$ is the supercell volume, $\bm{S}^{\alpha {\alpha'}}_{iR,i'R'} = \frac{1}{i}(\bm{R}+\bm{r}_i-\bm{R'}-\bm{r}_{i'}) D^{\alpha {\alpha'}}_{iR,i'R'}$ is the Hardy heat current operator \cite{hardyEnergyFluxOperatorLattice1963} and $W_\text{ph} = \frac{3}{\pi^2} \left( \frac{\hbar \omega}{k_B T} \right)^2 \left( - \frac{\partial f_B}{\partial \omega} \right)$ acts as a normalized half-window of width $\approx 2 k_B T$ centered on $\omega = 0$, with $f_B$ the Bose-Einstein distribution. In this approach, the phonon-phonon interactions are taken into account as a VCA frequency-dependent inverse lifetime $\Gamma_\text{3ph}$ playing the role of an inelastic damping in the phonon Green's function. The quantity $\frac{\pi k_B ^2 T}{3 \hbar \Omega} \text{Tr} \left[ \text{Im} \matr{G} \; \matr{S} \; \text{Im} \matr{G} \;  \matr{S} \right]$ can be defined as the phonon transport distribution function (TDF) $\Sigma_\text{ph}(\omega)$, in analogy with the case of thermoelectric transport (see equation~(11) in Ref.~\onlinecite{scheidemantelTransportCoefficientsFirstprinciples2003}, where $-\partial f_0/\partial \epsilon$ is the normalized window in the electronic case). 

\section{The Chebyshev polynomials Green's function method}
\label{appendix_CPGF}

The Chebyshev polynomials Green's function (CPGF) method has been reviewed in Refs.~\cite{ferreiraCriticalDelocalizationChiral2015,weisseKernelPolynomialMethod2006} for electrons, and has been adapted in Refs.~\onlinecite{bouzerarDrasticEffectsVacancies2020,thebaudSuccessBreakdownTmatrix2020} for phonons. Here, we give a brief presentation of the approach. The phonon Green's function of a large disordered supercell is expanded on the Chebyshev polynomial basis:
\begin{equation}
\label{CPGF_expansion}
\matr{G}(\bar{\omega}) = \sum^\infty_{n=0} g_n((\bar{\omega} + i \bar{\eta})^2) T_n(\bar{\matr{D}})
\end{equation}
where the bar indicates that the spectrum has been rescaled to $[-1,1]$, the $g_n(z)$ are known complex functions:
\begin{equation}
g_n(z) = -i (2-\delta_{n,0}) \frac{(z - i\sqrt{1-z^2})^n}{\sqrt{1-z^2}}
\end{equation}
and the $T_n(\bar{\matr{D}})$ are Chebyshev polynomials evaluated for the dynamical matrix, that follow the recursion relation $T_{n+1}(\bar{\matr{D}}) = 2 \bar{\matr{D}} T_{n}(\bar{\matr{D}}) - T_{n-1}(\bar{\matr{D}})$ with $T_{1}(\bar{\matr{D}}) = \bar{\matr{D}}$ and $T_{0}(\bar{\matr{D}}) = 1$. Equality~\eqref{CPGF_expansion} comes from the identity 
\begin{equation}
e^{-izt} = \sum^\infty_{n=0} \frac{2 i^{-n}}{1 + \delta_{n,0}} J_n(t) T_n(z)
\end{equation}
for $|z|<1$ with $J_n(t)$ the Bessel function of order $n$ (see eqs. (5) through (9) in the supplementary material of Ref.~\onlinecite{ferreiraCriticalDelocalizationChiral2015}).

Since the spectral function for the Bloch mode $(\bm{q}j)$ is given by eq.~\eqref{def_spectral_function}, the quantities to be calculated are the so-called moments $\mu_{n,\bm{q}j}$:
\begin{equation}
\mu_{n,\bm{q}j} = \bra{E_{\bm{q}j}} \sqrt{\frac{\matr{M}_\text{vc}}{\matr{M}}} T_n(\bar{\matr{D}}) \sqrt{\frac{\matr{M}_\text{vc}}{\matr{M}}} \ket{E_{\bm{q}j}},
\end{equation}
which are computed using the recursion relation between the $T_{n}(\bar{\matr{D}})$. Once this is done, the spectral function can be obtained at any $\omega$ at virtually no computational cost. The number of moments necessary for the sum~\eqref{CPGF_expansion} to converge is roughly equal to $1/2\bar{\omega}\bar{\eta}$. Because $\bar{\eta}$ is an artifical broadening and should be smaller than the disorder-induced spectral linewidth, probing modes closer and closer to $\Gamma$ requires more and more polynomials to be included. 

For the phonon DOS, the trace in eq.~\eqref{def_dos} is evaluated by a stochastic method detailed in Ref.~\onlinecite{weisseKernelPolynomialMethod2006}. We define $N_r$ random vectors $\ket{r} = \sum_{\alpha i \bm{R}} e^{i \varphi^{\alpha}_{i\bm{R}}} \ket {\alpha i \bm{R}}$ with $\varphi^{\alpha}_{i\bm{R}}$ a random phase uniformly distributed in the interval $[0,2 \pi]$. The DOS is then calculated as
\begin{equation}
\rho(\omega) \approx -\frac{2 \omega}{\pi} \frac{1}{N_r} \sum_r \bra{r} \text{Im}  \matr{G}(\omega) \ket{r}.
\end{equation}
The sum over the different random vectors acts as an averaging procedure for the random phases, so only the diagonal terms survive, yielding a trace. In the same way as for the spectral function, we compute the moments 
\begin{equation}
\mu_{n,qj} =  \frac{1}{N_r} \sum_r \bra{r} T_n(\bar{\matr{D}})  \ket{r},
\end{equation}
and then the DOS can be easily obtained on the whole spectrum.

To calculate the phonon TDF from eq.~\eqref{kubo_kappa}, we use the one-shot procedure described in Ref.~\onlinecite{ferreiraCriticalDelocalizationChiral2015}:
\begin{equation}
\Sigma_\text{ph}(\omega) =  \frac{\pi k_B ^2 T}{3 \hbar \Omega} \frac{1}{N_r} \sum_r \braket{\varphi_-^{(r)} | \varphi_+^{(r)}}
\end{equation}
with $\ket{\varphi_+^{(r)}} = \text{Im} \matr{G}(\omega + i \frac{\Gamma_\text{3ph}(\omega)}{2}) \matr{S} \ket{r}$ and $\ket{\varphi_-^{(r)}} = \matr{S} \, \text{Im} \matr{G}(\omega + i \frac{\Gamma_\text{3ph}(\omega)}{2}) \ket{r}$. The vectors $\ket{\varphi_+^{(r)}}$ and $\ket{\varphi_-^{(r)}}$ are calculated iteratively. Unlike for the DOS and spectral function, we do not compute the moments of the TDF. Consequently, this procedure has to be repeated for every frequency $\omega$, but it is much less demanding in memory than the alternative of computing the moments.   

In order to evaluate the DOS, spectral function and TDF on very large supercells ($\sim$ \num{e7} atoms), we exploit the sparsity of the dynamical matrix in the real-space basis of atomic displacements by storing it in the compressed sparse row representation. Thus, both the memory requirement and computation time for the iterative calculation of the moments and vectors $\ket{\varphi_+^{(r)}}$ and $\ket{\varphi_-^{(r)}}$ scale linearly with the supercell size $N_c$ as opposed to the respectively quadratic and cubic scaling of exact diagonalization methods. Moreover, the Chebyshev expansion converges exponentially for a given imaginary part $\bar{\eta}$ \cite{vijayScatteringBoundStates2004}. These properties lead to reasonable memory requirements and calculation times even for systems of tens of millions of atoms.

For the CPGF calculation of the DOS, we consider supercells of 125000 atoms with 50000 moments and 5 random vectors to evaluate the trace. For the spectral function, we build supercells of up to \num{5e6} atoms ($120 \times 120 \times 120$ unit cells) and include up to \num{3e5} moments in the expansion of the Green's function. For such system sizes, the quantities are self-averaged and it is sufficient to consider only one disorder configuration. For reasonably defined quasiparticle peaks in the spectral function, we extract the inverse phonon lifetimes by evaluating the full-width at half maximum (FWHM) of the  peaks. This is done by fitting the peaks by a Lorentzian or the product of a Lorentzian and a linear function for acoustic peaks displaying a high degree of asymmetry. A spline interpolation is then performed to calculate the FWHM. 

For the CPGF calculation of the TDF, we build supercells of up to \num{16e6} atoms ($1500 \times 60 \times 60$ unit cells) and include up to 8000 moments. We use 4 random vectors per frequency for the trace evaluation, and we estimate the error associated with this procedure by the standard variance formula (95\% confidence interval):
\begin{equation}
\Delta \Sigma_\text{ph} = 2\sqrt{\frac{1}{N_r(N_r-1)} \sum_r (\Sigma_\text{ph}^{(r)} - \Sigma_\text{ph})^2} 
\end{equation}
where $\Sigma_\text{ph}^{(r)}$ is the TDF value from random vector $r$ and $\Sigma_\text{ph} = \frac{1}{N_r} \sum_r \Sigma_\text{ph}^{(r)}$.

\section{Higher-order perturbative expansion of the phonon self-energy}
\label{appendix_diagrams}

Some insights on the success and failure of the FGR can be obtained from a standard diagrammatic expansion of the phonon Green's function with respect to mass disorder. We will briefly introduce the theory here, but the interested reader can consult, e.g., section III of Ref.~\onlinecite{elliottTheoryPropertiesRandomly1974}. The spectral function for VCA mode $(\bm{q}j)$ can be expressed as  
\begin{equation}
\label{def_spectral_function_diagram}
A_j(\bm{q},\omega) = -\frac{2 \omega}{\pi}\text{Im} \bra{E_{\bm{q}j}} \tilde{\matr{G}}(\omega) \ket{E_{\bm{q}j}}
\end{equation}
with the mass-renormalized Green's function
\begin{align}
\label{def_mass_renormalized_green}
\tilde{\matr{G}}(\omega) & = \sqrt{\frac{\matr{M}_\text{vc}}{\matr{M}}} \matr{G}(\omega) \nonumber \sqrt{\frac{\matr{M}_\text{vc}}{\matr{M}}}  \\
& = \frac{1}{(\omega + i \eta)^2 - \matr{D}_\text{vc} - \matr{P}} .
\end{align}
The mass perturbation $\matr{P} = - \frac{\Delta \matr{M}}{\matr{M}_\text{vc}} \omega^2$ can be decomposed as
\begin{equation}
\label{mass_perturbation}
\matr{P} = \sum_{\bm{R}} \matr{P}_{\bm{R}} = \sum_{\bm{R}} (- \omega^2) \frac{\Delta m_{\bm{R}}}{m_\text{vc}} \sum_\alpha \ket{\alpha 1 \bm{R}} \bra{\alpha 1 \bm{R}} 
\end{equation}
where $m_\text{vc} = (m_\text{Si} + m_\text{Sn})/2$ is the mass of the `virtual' Si/Sn atom in the VCA, $\Delta m_{\bm{R}} = m_\text{Si} - m_\text{vc}$ if unit cell $\bm{R}$ contains a Si atom or $m_\text{Sn} - m_\text{vc}$ otherwise, and $\ket{\alpha 1 \bm{R}}$ denotes the degree of freedom along $\alpha$ of the Si/Sn atom in unit cell $\bm{R}$.

\begin{figure}
\centering
\includegraphics[width=0.6\columnwidth]{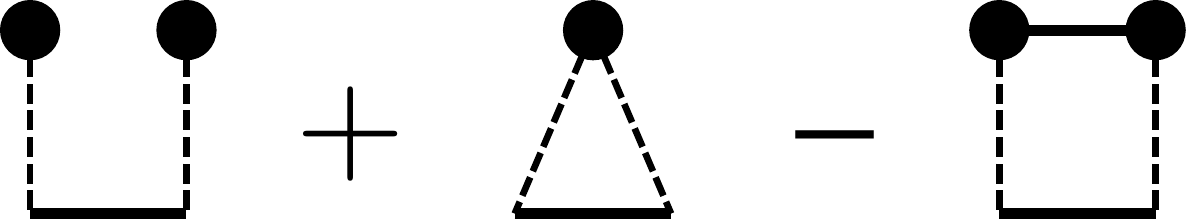}
\caption{The Feynman diagrams representing the three terms in eq.~\eqref{diagrams_order2}. See text.}
\label{fig7}
\end{figure}

Performing a Taylor expansion of expression~\eqref{def_mass_renormalized_green} with respect to the mass perturbation and taking the configuration average yields
\begin{align}
\langle \tilde{\matr{G}} \rangle_c & = \matr{G}_\text{vc} + \sum_{\bm{R}} \matr{G}_\text{vc} \langle \matr{P}_{\bm{R}} \rangle_c \matr{G}_\text{vc} \\
& + \sum_{\bm{R},\bm{R'}}  \matr{G}_\text{vc} \langle \matr{P}_{\bm{R}} \matr{G}_\text{vc} \matr{P}_{\bm{R'}} \rangle_c \matr{G}_\text{vc} + ... \nonumber
\end{align}
with the VCA Green's function $\matr{G}_\text{vc} =  \frac{1}{(\omega + i \eta)^2 - \matr{D}_\text{vc}}$. Assuming that the atomic occupations are uncorrelated from one unit cell to another, the last term in this equation can be decomposed:
\begin{multline}
\label{diagrams_order2}
 \sum_{\bm{R},\bm{R'}} \langle \matr{P}_{\bm{R}} \matr{G}_\text{vc} \matr{P}_{\bm{R'}} \rangle_c \\   = \sum_{\bm{R} \neq \bm{R'}} \langle \matr{P}_{\bm{R}}  \rangle_c \matr{G}_\text{vc} \langle \matr{P}_{\bm{R'}} \rangle_c  + \sum_{\bm{R}} \langle \matr{P}_{\bm{R}} \matr{G}_\text{vc} \matr{P}_{\bm{R}} \rangle_c \\
 = \sum_{\bm{R}} \langle \matr{P}_{\bm{R}}  \rangle_c \matr{G}_\text{vc} \sum_{\bm{R'}}\langle \matr{P}_{\bm{R'}} \rangle_c  +  \sum_{\bm{R}} \langle \matr{P}_{\bm{R}} \matr{G}_\text{vc} \matr{P}_{\bm{R}} \rangle_c \\
 - \sum_{\bm{R}} \langle \matr{P}_{\bm{R}}  \rangle_c \matr{G}_\text{vc} \langle \matr{P}_{\bm{R}} \rangle_c .
\end{multline}
The three terms in the last expression can be represented diagrammatically, as shown in Fig.~\ref{fig7}. Each lower horizontal line represents the VCA Green's function and each dashed line a perturbation operator on unit cell $\bm{R}$. Each dot represents a configuration average that must be taken over the associated perturbation operators. The upper horizontal lines connecting different dots indicate that the dots share the same position $\bm{R}$, and a sum over $\bm{R}$ is implied for each group of connected dots. Notice that the first term of the last expression is made up of two independent operators connected by a VCA Green's function, i.e., its diagram can be cut in two by slicing one lower horizontal line. It is called a reducible diagram, while the other two terms are called irreducible diagrams. The very last term is part of a set of Feynman diagrams called multiple occupancy corrections (MOC), which ensure that the $\bm{R} = \bm{R'}$ cases are not double counted. They can be neglected in the case of dilute impurities, but are very important for alloys (see below).  

\begin{figure}
\centering
\includegraphics[width=1.0\columnwidth]{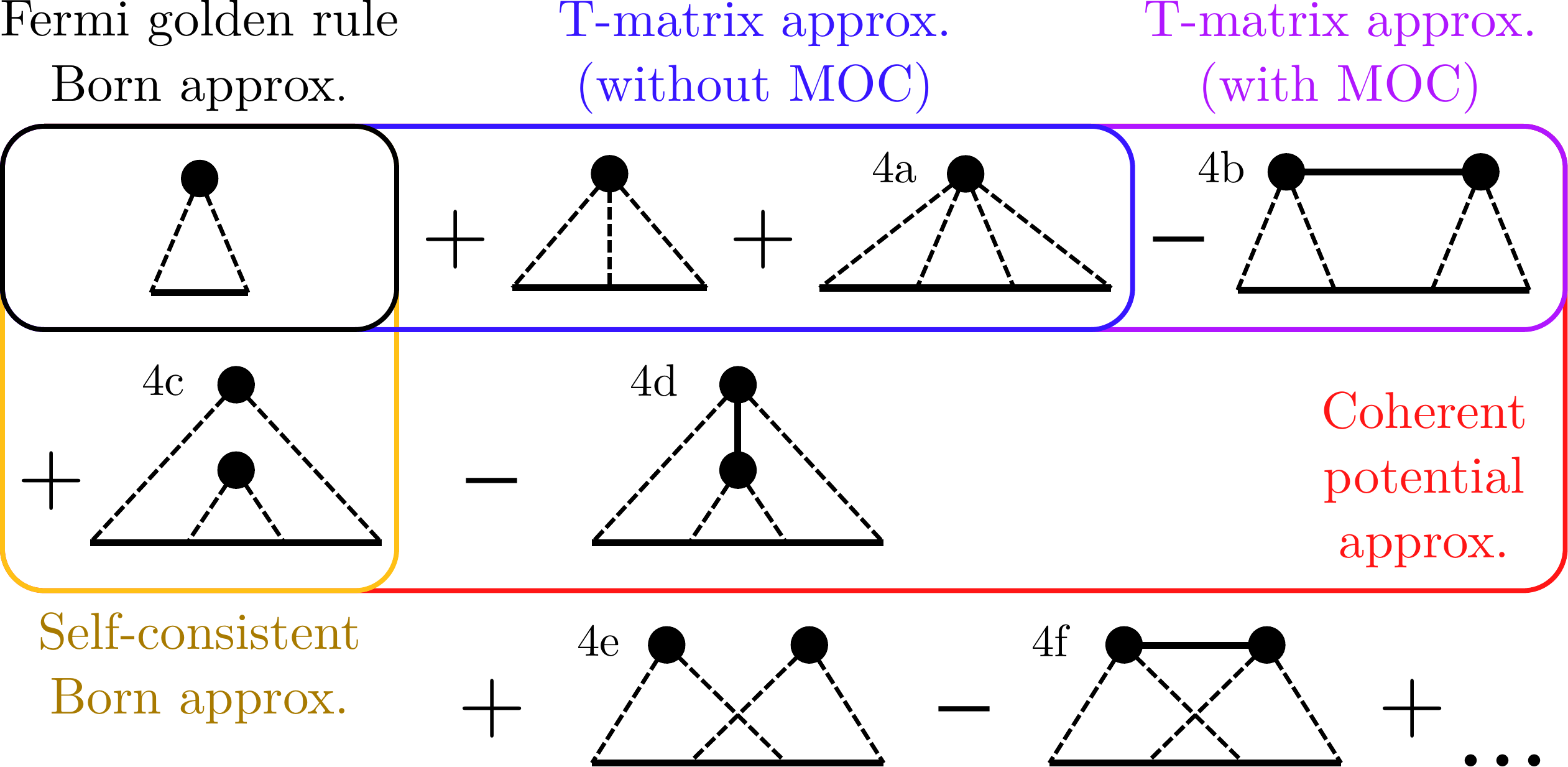}
\caption{The Feynman diagrams appearing in the diagrammatic expansion for $2 \omega \matr{\Pi}$ in a binary alloy up to fourth order.}
\label{fig8}
\end{figure}

The sum of the irreducible diagrams defines the phonon self-energy operator $\matr{\Pi}(\omega)$ through a Dyson equation for the Green's function:
\begin{align}
\langle \tilde{\matr{G}} \rangle_c & = \matr{G}_\text{vc} + \matr{G}_\text{vc} 2 \omega \matr{\Pi} \matr{G}_\text{vc} + \matr{G}_\text{vc} 2 \omega \matr{\Pi} \matr{G}_\text{vc}  2 \omega \matr{\Pi} \matr{G}_\text{vc} + ... \nonumber \\
& = \frac{1}{\omega^2 - \matr{D}_\text{vc} - 2 \omega \matr{\Pi}}
\end{align}
where the small imaginary part $\eta$ can now be removed since the self-energy is a complex quantity. For weak disorder, the phonon-disorder scattering rate of the VCA mode $(\bm{q}j)$ is given by $\Gamma_{\bm{q}j} = - 2 \text{Im} \braket{E_{\bm{q}j} | \matr{\Pi}(\omega_{\bm{q}j}) | E_{\bm{q}j}}$ (see section III D of Ref.~\onlinecite{elliottTheoryPropertiesRandomly1974} and section 7.2 of Ref.~\onlinecite{economouGreenFunctionsQuantum1979}). We show in Fig.~\ref{fig8} the self-energy diagrams up to fourth order in the case of a binary mass-disordered alloy, assuming a VCA reference crystal such that diagrams featuring a dot connected to only one dashed line vanish. The fourth-order diagrams are labeled from 4a to 4f. The second-order diagram is simply the FGR (also called the Born approximation). The third and fourth-order diagram 4a appear in the T-matrix approximation without MOC, which is exact in the limit of dilute impurities and has been used in thermal transport calculations in the literature \cite{polancoDefectlimitedThermalConductivity2020,wangInitioPhononScattering2017,dongreResonantPhononScattering2018}.   Diagram 4b is part of the T-matrix approximation that includes MOC, which is well-known as the average T-matrix approximation (ATA) in the early literature on electron and phonon scattering in alloys \cite{elliottTheoryPropertiesRandomly1974,schwartzComparisonAveragetMatrixCoherentPotential1971}.   Diagram 4c is nested, i.e., it can be obtained by inserting the FGR diagram in the internal VCA Green's function of that same diagram. As such, they are part of the self-consistent Born approximation (SCBA) which demands that the Green's function obtained from the FGR self-energy matches the internal Green's function in the FGR diagram. Diagram 4d is part of the coherent potential approximation (CPA), which demands the same thing from the ATA self-energy and thus includes all nested diagrams generated from the ATA. Finally, diagrams 4e and 4f are not part of any standard approximation scheme as they feature crossed scatterings. We stress that the CPGF method, being non-perturbative, includes all the diagrams to infinite order.   

We now turn to the issue of evaluating at which frequency the FGR breaks down. Because each dashed scattering line brings a factor $\omega^2$, higher-order diagrams correspond to higher powers of $\omega$, and become negligible compared to the FGR term below a certain frequency. In the case of a maximally disordered binary alloy, such as Mg$_2$Si$_{0.5}$Sn$_{0.5}$, the third-order diagram vanishes and diagrams 4a, 4b, 4d and 4f turn out to be equal, such that the ATA gives the same result as the FGR (this is actually true at all orders). Therefore, we restrict our analysis to the ratio of the fourth-order contribution to the scattering rate with respect to the FGR scattering rate:
\begin{equation}
\frac{\Gamma^\text{4}}{\Gamma^\text{FGR}} = \frac{\Gamma^\text{4c}}{\Gamma^\text{FGR}} + \frac{\Gamma^\text{4e}}{\Gamma^\text{FGR}} - 2 \frac{\Gamma^\text{4a}}{\Gamma^\text{FGR}} .
\end{equation}
To evaluate the relative magnitude of these terms, we have considered a simple Debye model with a relative mass perturbation $\pm \delta m$ on each atom and a Debye frequency $\omega_D$. An analytical calculation of the dominant contributions from each term in the regime $\omega \ll \omega_D$ yields
\begin{equation}
\frac{\Gamma^\text{4}_\text{Deb}}{\Gamma^\text{FGR}_\text{Deb}} = \frac{3}{2} \delta m^2 \frac{\omega^2}{\omega_D^2} + \frac{3\pi^2}{32} \delta m^2 \frac{\omega^3}{\omega_D^3} - 54  \delta m^2 \frac{\omega^4}{\omega_D^4}.
\end{equation}
The last term dominates (except below $\approx 0.15 \omega_D$ where these corrections are negligible), therefore the breakdown of FGR is initiated by the onset of the term $- 2 \, \Gamma^\text{4a}$, which is negative at low frequencies. This explains the somewhat counterintuitive result that the FGR tends to \textit{overestimate} the scattering rate.  We stress that the dominant correction to the FGR, $- 2 \, \Gamma^\text{4a}$, comes from four different diagrams in the perturbative expansion of the self-energy, including three MOC diagrams, one of which cannot be obtained by any standard approximation scheme such as the T-matrix approximation or the CPA.

\begin{figure}
\centering
\includegraphics[width=1.0\columnwidth]{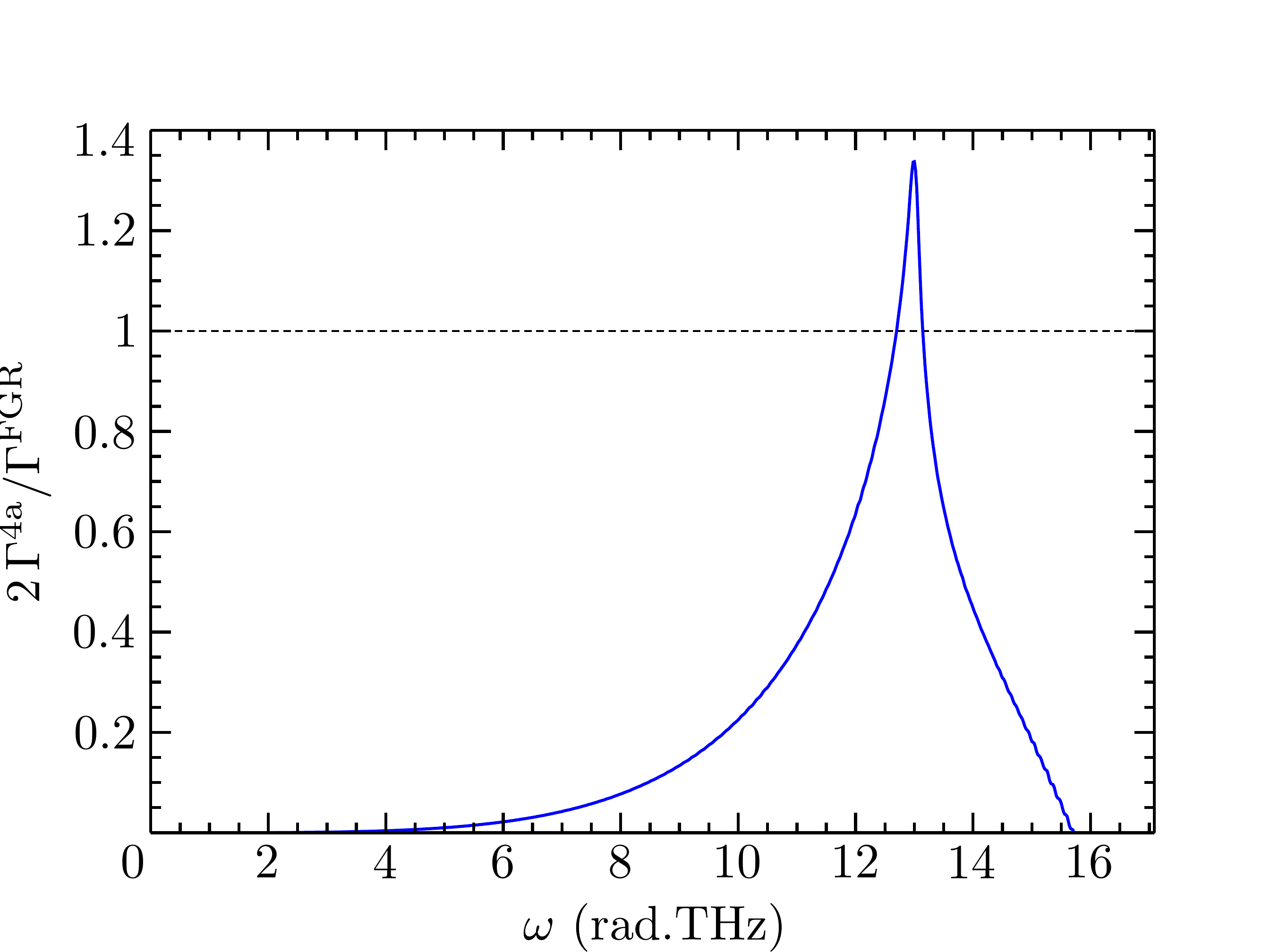}
\caption{Ratio of the dominant fourth-order contribution to the scattering rate with respect to the FGR scattering rate as a function of frequency for Mg$_2$Si$_{0.5}$Sn$_{0.5}$.}
\label{fig9}
\end{figure}

In Mg$_2$Si$_{0.5}$Sn$_{0.5}$, the scattering rate ratio can be written for mode $(\bm{q}j)$ as
\begin{equation}
\label{srate_ratio_mg2sisn}
 2 \frac{\Gamma_{\bm{q}j}^\text{4a}}{\Gamma_{\bm{q}j}^\text{FGR}} = 2 \delta m^2 \omega_{\bm{q}j}^4 \frac{\text{Im}(\vec{e}_{1,\bm{q}j}^{\; *} \cdot {\cal G}(\omega_{\bm{q}j})^3 \cdot \vec{e}_{1,\bm{q}j})}{\text{Im}(\vec{e}_{1,\bm{q}j}^{\; *} \cdot {\cal G}(\omega_{\bm{q}j}) \cdot \vec{e}_{1,\bm{q}j})} 
\end{equation}
where $\vec{e}_{1,\bm{q}j}$ is the polarisation vector on the Si/Sn atom of the unit cell and ${\cal G}(\omega)$ is a tensor obtained by restricting the VCA Green's function to the degrees of freedom of the Si/Sn atom of the central unit cell:
\begin{equation}
\label{green_tensor}
{\cal G}_{\alpha \beta}(\omega) = \frac{1}{N_c}\sum_{\bm{q}j} \frac{e^\alpha_{1,\bm{q}j} \, e^{\beta \; *}_{1,\bm{q}j}}{(\omega + i \eta)^2 - \omega^2_{\bm{q}j}}.
\end{equation}
Due to the cubic symmetry of Mg$_2$Si$_{0.5}$Sn$_{0.5}$, ${\cal G}$ is actually a scalar matrix in this case and the ratio~\eqref{srate_ratio_mg2sisn} depends only on the frequency, as shown in Fig.~\ref{fig9}. The fourth-order term starts becoming non-negligible around \SIrange{10}{11}{rad.THz}, leading to the FGR being slightly overestimated as can be seen in Fig.~\ref{fig2}. This effect is somewhat compensated by the fact that the VH singularity is displaced to lower frequencies in the disordered system (see Fig.~\ref{fig1}). However, the VH singularity of the VCA spectrum at \SI{13}{rad.THz} leads to a peak in the fourth-order term such that the FGR breaks down completely at this frequency. Thus, using the standard diagrammatic expansion with respect to mass perturbations, it is possible to determine \textit{a priori} the frequency regime where the FGR is valid in mass-disordered alloys. This can be done by evaluating expressions such as \eqref{srate_ratio_mg2sisn} and \eqref{green_tensor}, which require only the VCA spectrum routinely obtained from lattice dynamics calculations. For completeness, we give them for the more general formula Mg$_2$Si$_{1-x}$Sn$_x$, with $m_\text{vc} = (1-x) m_\text{Si} + x m_\text{Sn}$, $\delta m_\text{Si} = (m_\text{Si} - m_\text{vc}) / m_\text{vc}$ and $\delta m_\text{Sn} = (m_\text{Sn} - m_\text{vc}) / m_\text{vc}$. In the fourth-order ratio \eqref{srate_ratio_mg2sisn}, the prefactor $2\delta m^2$ should be replaced by $\delta m_\text{Si} \delta m_\text{Sn} [3 - \frac{x^2}{1-x} - \frac{(1-x)^2}{x}]$. The ratio between the third-order contribution and the FGR contribution is in general non-zero: the prefactor $2\delta m^2 \omega_{\bm{q}j}^4$ should be replaced by $[x\delta m_\text{Si} + (1-x) \delta m_\text{Sn}]  \omega_{\bm{q}j}^2$ and ${\cal G}^3$ by ${\cal G}^2$ in the numerator. We stress that the VCA frequencies should be recalculated for each $x$ value.

\begin{footnotesize}
\bibliographystyle{unsrt} 
\bibliography{../../biblio/biblio_simon.bib}

\end{footnotesize}

\end{document}